\documentclass[10pt, aps, prb, twocolumn, floatfix, superscriptaddress]{revtex4-2}

\usepackage{amsmath,amssymb}
\usepackage{graphicx}
\usepackage{bm}
\usepackage{mathtools}
\usepackage{booktabs}
\usepackage{braket}
\usepackage[T1]{fontenc}
\usepackage[hidelinks,bookmarks=false]{hyperref}
\usepackage[capitalise,nameinlink]{cleveref}
\crefname{figure}{Extended Data Fig}{Extended Data Fig}

\makeatletter
\def\fnum@figure{\textbf{Fig.~\thefigure}}
\makeatother

\begin{document}

\title{Simultaneous operation of an 18-qubit modular array in germanium}
\author{Jurgen~J.~Dijkema}
\altaffiliation{These authors contributed equally to this work}
\author{Xin~Zhang}
\altaffiliation{These authors contributed equally to this work}
\affiliation{Groove Quantum B.V., Lorentzweg 1, 2628 CJ Delft, Netherlands}
\author{Achilleas~Bardakas}
\author{Daniel~Bouman}
\author{Alice~Cuzzocrea}
\author{David~van~Driel}
\author{Davide~Girardi}
\author{Lucas~E.A.~Stehouwer}
\affiliation{Groove Quantum B.V., Lorentzweg 1, 2628 CJ Delft, Netherlands}
\author{Giordano~Scappucci}
\affiliation{QuTech and Kavli Institute of Nanoscience, Delft University of Technology, Lorentzweg 1, 2628 CJ Delft, Netherlands}
\author{Anne-Marije~J.~Zwerver}
\author{Nico~W.~Hendrickx}
\affiliation{Groove Quantum B.V., Lorentzweg 1, 2628 CJ Delft, Netherlands}

\date{\today}

\begin{abstract}
Utility-scale quantum computing requires the integration and operation of a large-scale qubit register. Semiconductor spin qubits are a primary candidate for this, due to the prospects of building integrated hybrid quantum-classical architectures. However, scaling spin-qubit systems while preserving performance and control has remained a challenge. Here, we demonstrate the operation of an 18-qubit array in germanium based on an extendable $2\times N$ architecture. We achieve simultaneous initialization, control, and readout across the entire array, enabled by parallel operation of modular unit cells. Across the array, we achieve average and median single-qubit gate fidelities of $99.8$\% and $99.9$\%, respectively. Finally, we characterize the nearest-neighbor exchange couplings throughout the device and implement high-quality controlled-Z gates to generate a three-qubit Greenberger-Horne-Zeilinger (GHZ) state. These results demonstrate that spin-qubit arrays can be scaled while maintaining high-fidelity operation and establish a modular, extendable architecture for planar semiconductor quantum processors.
\end{abstract}

\maketitle

\section{Introduction}
The realization of utility-scale quantum computing depends on the development of hardware architectures that can scale to a large number of qubits while efficiently managing the associated wiring and control overhead~\cite{mohseni2024build}. Semiconductor-based quantum technologies are particularly promising in this regard, as they build directly on the mature infrastructure of the semiconductor industry~\cite{zwerver2022qubits, neyens2024probing, steinacker2025industry}. Leveraging advanced manufacturing and packaging technologies can enable the integration of quantum hardware with classical control electronics~\cite{xue2021cmos,bartee2025spin}, thereby realizing the required quantum-classical architectures that mitigate the wiring bottleneck~\cite{vandersypen2017interfacing, veldhorst2017silicon, franke2019rent, gonzalez2021scaling}. 

Rapid progress in small-scale devices has enabled high-fidelity qubit operations~\cite{stano2022review, xue2022quantum, mills2022two, mkadzik2022precision, huang2024high, wu2025simultaneous}, initial demonstrations of error-correction codes~\cite{takeda2022quantum, van2022phase, undseth2026parity, zhang2026quantum, zhang2026universal}, and remote entanglement through qubit shuttling~\cite{de2025high, matsumoto2025two, ademi2025distributing}. In parallel, array sizes have been scaled to 6–12 qubits in one-dimensional silicon devices~\cite{philips2022universal, george2025intel12qb, nickl2025eight, edlbauer202511} and 4–10 qubits in two-dimensional germanium architectures~\cite{hendrickx2021four, zhang2025universal, john2024two}. However, these demonstrations have remained largely confined to individual, sequential qubit operations and often rely on bespoke, non-modular device layouts. Consequently, an experimental demonstration of a modular, extensible unit-cell architecture capable of supporting simultaneous high-fidelity control and parallelized readout across an integrated two-dimensional grid has been lacking.

Here, we present an 18-qubit germanium quantum processing unit (QPU) based on a $2\times N$ architecture that is extendable to arbitrary length. The device is composed of modular six-qubit unit cells with dedicated charge sensors and achieves single-qubit gate fidelities above 99\% across the full array. Furthermore, we demonstrate that different unit cells can be initialized, operated, and read out in parallel, ensuring that the overhead associated with state preparation and measurement (SPAM) does not increase with system size. These results constitute the largest spin-qubit array demonstrated to date and establish a blueprint for scalable two-dimensional semiconductor quantum processors.

\begin{figure*}[htbp] 
\center{\includegraphics[width=\linewidth]{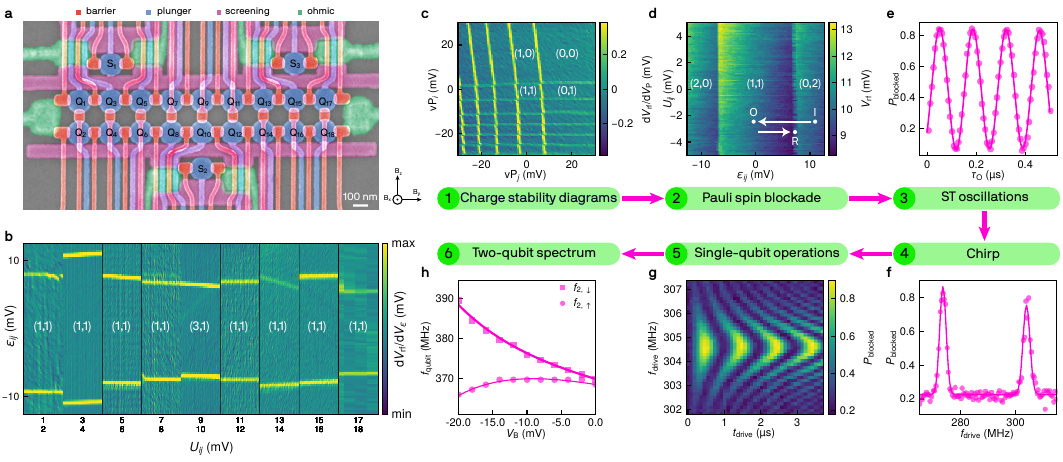}}
\caption{\textbf{Device and measurement protocol of the 18-qubit modular array.} \textbf{a}, False-colored scanning electron microscope image of the 18-qubit device. Ohmic contacts, screening gates, plunger gates, and barrier gates are indicated in green, magenta, blue and red, respectively. The 18 qubits, located below the plunger gates, are labeled Q$_{1}$-Q$_{18}$. Three charge sensors S$_i$ ($i$ the unit cell index) are positioned above (S$_1$, S$_3$) and below (S$_2$) the array. The reference frame of the external magnetic field \text{$B_{x,y,z}$} is indicated by the black arrows.
\textbf{b}, Few-hole charge stability diagram in the operational regime for all vertical double quantum dot (DQD) pairs Q$_i$Q$_j$, shown as a function of the detuning ($\epsilon_{ij}$) and on-site ($U_{ij}$) energy.
\textbf{c-h}, Tuning procedure to characterize the spin qubits.
\textbf{c}, Charge stability diagram of a representative DQD, measured as a function of the virtual plunger gates voltages $\text{vP}_i$ and $\text{vP}_j$.
\textbf{d}, The charge stability diagram of the two-hole regime shows the relevant charge transitions, with Pauli spin blockade (PSB) appearing as a latched signal near the (1,1)-(0,2) transition, when sweeping towards positive detuning. A typical pulse sequence, including initialization (I), operation (O) and readout (R), is indicated.
\textbf{e}, Singlet-triplet ($\text{S}-\text{T}_0$) oscillations measured in the blocked state probability $P_\text{blocked}$ as a function of the dwell time at the operation point $\tau_\text{O}$.
\textbf{f}, Spin resonance spectroscopy using a chirped microwave pulse, showing two resonance peaks corresponding to the two spins in the DQD.
\textbf{g}, Rabi chevron pattern showing the spin-flip probability as a function of microwave pulse duration ($t_\text{drive}$) and drive frequency ($f_\text{drive}$).
\textbf{h}, Exchange-coupling spectroscopy, where the splitting of the target qubit resonance frequency $f_\text{qubit}$ as a function of the applied barrier gate voltage $V_\text{B}$, with the control qubit initialized in the $\ket{\downarrow}$ (squares) or $\ket{\uparrow}$ (circles) state.}
\label{fig:1}
\end{figure*}

\begin{figure*}[htbp] 
\center{\includegraphics[width=\linewidth]{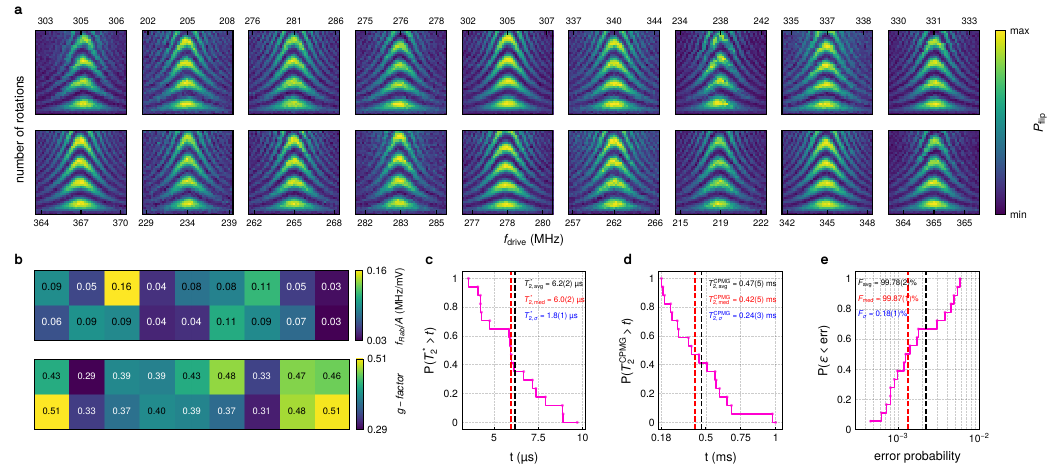}}
\caption{\textbf{Statistical analysis of single-qubit coherence and gate fidelities across the array.} \textbf{a}, Chevron pattern measurements for all 18 spin-qubits, showing four full rotations as a function of the drive frequency ($f_{\rm drive}$). All measurements are performed sequentially without changing the dc voltage settings of the device. \textbf{b}, Color scale mapping of the qubit driving efficiency ($f_\text{Rabi}/A$, top panel) and $g$-factors (bottom panel) for all qubits. \textbf{c-e}, Empirical cumulative distribution function (ECDF) of $T_2^*$ (\textbf{c}), $T_2^\text{CPMG}$ (\textbf{d}), and the single qubit gate error probability (\textbf{e}) of all 18-qubits, respectively. The black and red dashed lines show the average and median values, respectively. The mean, median, and standard deviation of the distribution are summarized in the top right corner.}
\label{fig:2}
\end{figure*}

\section{Device and measurement protocol}
The 18-qubit device, as shown in Fig.~\ref{fig:1}a, is defined in a Ge/SiGe heterostructure~\cite{lodari2021low, scappucci2021germanium} and consists of three repeated 2x3-qubit unit cells. The unit-cell design features a locally one-dimensional gate fan-out, enabling straightforward extension of the architecture to larger arrays (see \cref{fig:2xNscheme}). Each unit cell contains a dedicated charge sensor that can read out the charge state of all quantum dots within the cell. The external magnetic field orientation is chosen such that all qubits operate near the in-plane sweet spot of minimal hyperfine interaction~(\cref{fig:Bx_field_sweep})~\cite{fischer2008spin, hendrickx2024sweet,yu2025optimising}, at $B_x$=0.83~mT, $B_y$=50~mT, and $B_z$=10~mT. The qubits are labeled Q$_1$-Q$_{18}$, and the three charge sensors are labeled S$_1$, S$_2$, and S$_3$, respectively.

Using the three charge sensors together, we can resolve the charge configuration of the entire quantum dot array. Fig.~\ref{fig:1}b presents the charge stability diagram (CSD) for all vertical double quantum dot (DQD) pairs as a function of their detuning $\varepsilon_{ij}=(\text{vP}_i-\text{vP}_j)/2$ and total energy $U_{ij}=(\text{vP}_i+\text{vP}_j)/2$, where $\text{vP}_i$ and $\text{vP}_j$ denote the virtualized plunger gate voltages~\cite{hensgens2017quantum} of quantum dots $i$ and $j$, respectively (details in Methods). For each DQD, except Q9,Q10, the charge configuration at $\varepsilon_{ij}=0$ corresponds to the $(1, 1)$ regime, where ($N_i$, $N_j$) denotes the hole occupation of quantum dots $i$ and $j$.

To characterize the single-qubit properties across the array, we follow the tuning procedure illustrated in Fig.~\ref{fig:1}c-h. First, individual CSDs are measured to obtain the full 18-quantum-dot charge configuration, with an example shown in Fig.~\ref{fig:1}c (further CSD measurements of all DQD pairs are provided in \cref{fig:all_DQD_plots}). Next, Pauli spin blockade (PSB) is identified at the relevant interdot crossings, as illustrated in \ref{fig:1}d for the (1,1)-(0,2) charge transition. The readout of the spin parity in the DQD is confirmed through the observation of S-T$_0$ oscillations (Fig.~\ref{fig:1}e) as a function of the duration of a diabatic detuning pulse across the anti-crossing. While PSB readout can be observed for both vertical and horizontal DQDs, we operate the device by performing a parity readout of the vertical DQDs (readout pairs, see Methods for details). Next, we identify the single-spin qubit resonance frequencies by applying chirped microwave pulses to the plunger gates and detecting spin resonance. This is illustrated in Fig.~\ref{fig:1}f, where two resonance lines correspond to single-spin resonances of the two spins in the respective DQD. We confirm coherent qubit control by measuring the blocked state probability as a function of microwave drive pulse durations $t_\text{drive}$ and drive frequency $f_\text{drive}$ (Fig.~\ref{fig:1}g). Finally, we characterize the tunability of the exchange interaction within the DQD through spectroscopy while applying a voltage pulse of varying depth $V_{\rm B}$ to the interdot barrier gate. A chirped microwave burst is applied to the target qubit while preparing the control qubit in either basis state. As shown in Fig.~\ref{fig:1}h, the resonance frequency of the target qubit clearly splits depending on the state of the control qubit and the depth of the barrier voltage pulse, providing clear evidence of a tuneable exchange coupling between the qubits.

\section{Single-qubit gate}
Following the measurement protocol described above, we calibrate single-qubit gates for all 18 qubits while maintaining the same gate voltage offsets, enabling operation of all qubits within a single device configuration. All qubits are operated at comparatively low frequency, improving qubit coherence~\cite{hendrickx2024sweet} and reducing control architecture complexity. Single qubit control is demonstrated through the observation of 18 Rabi chevron patterns shown in Fig.~\ref{fig:2}a, where the drive amplitude and pulse duration are calibrated to observe four full qubit rotations. The spatial distribution of qubit $g$-factors and driving efficiencies $f_\text{Rabi}/A$, with $A$ the driving amplitude, are shown in Fig.~\ref{fig:2}b (more details in \cref{fig:18QB_Rabi&Larmor_frequencies}). Both the measured $g$-factors and driving efficiencies are consistent with other reports in germanium~\cite{hendrickx2024sweet, john2024two, seidler2025spatial}. The observed spread of the $g$-factors across the devices likely arises from strain fluctuations that can be reduced by employing Ge/SiGe heterostructures grown on Ge substrates~\cite{stehouwer2025exploiting, john2024two}.

Next, we characterize qubit coherence by performing Ramsey and Carr-Purcell-Meiboom-Gill (CPMG) measurements to extract the dephasing and coherence times $T_2^*$ and $T_2^\text{CPMG}$. The distributions of phase coherence times across the array are shown as empirical cumulative distribution functions (ECDFs) in Fig.~\ref{fig:2}c and d (details in \cref{fig:T2star} and \cref{fig:T2CPMG}). We find an average $T_{2,\text{avg}}^*$ of $6.2(2)~\mu$s, with a standard deviation of $T_{2,\sigma}^*=1.8(1)$~$\mu$s. The average $T_2^\text{CPMG}$ is $0.47(5)$~ms, with a standard deviation of $T_2^\text{CPMG}=0.24(3)$ ~ms. The large difference between $T_2^*$ and $T_2^\text{CPMG}$ indicates that coherence is limited by low-frequency noise (\cref{fig:T2CPMG}). We note that due to a finite spread of qubit g-tensor orientations (see \cref{fig:Bx_field_sweep}), the external magnetic field cannot be aligned with all hyperfine sweet spots~\cite{hendrickx2024sweet}. As a result, further improvements in qubit coherence are expected when operating in isotopically purified germanium quantum wells, where hyperfine noise is strongly suppressed for all field orientations~\cite{fischer2008spin, hendrickx2024sweet}.

To quantify the single-qubit control fidelities, we perform randomized benchmarking on all the qubits in the array (Fig.~\ref{fig:2}e, see Methods and \cref{fig:RBs_diff} for more details). For all qubits, the extracted primary gate fidelity $F_\text{g}$ exceeds $99.4$\%, with an average fidelity of $F_{\text{g},\mu}=99.8$\%. To suppress off-resonant driving of neighboring qubits, we employ Tukey-shaped microwave pulses, which reduce spectral leakage and mitigate crosstalk. Further improvements in control fidelity are expected by applying even more optimized pulse shaping~\cite{rimbach2023simple, wu2025simultaneous}. 

\begin{figure*}[htbp] 
\center{\includegraphics[width=\linewidth]{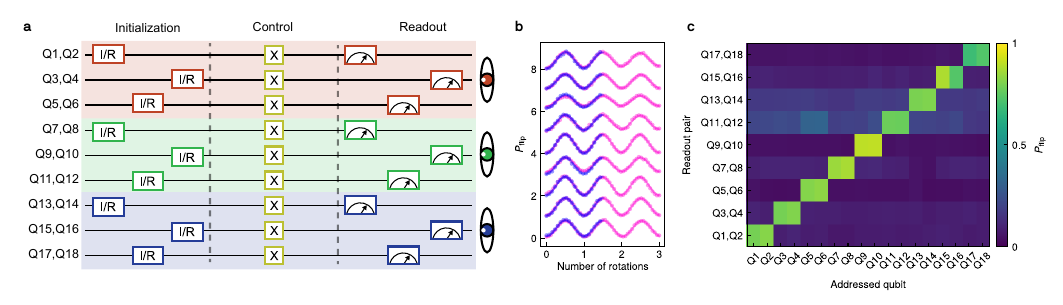}}
\caption{\textbf{Multi-qubit initialization, control and readout.} \textbf{a}, Pulse sequence used for multi-qubit initialization, control, and readout. The horizontal lines correspond to the vertical pairs of the array. The shaded regions indicate the three unit cells. Within each unit cell, initialization and reference operations (I/R) are executed sequentially across the three pairs, while the same sequence is performed in parallel across all unit cells. During the control stage, X gates are applied to individual qubits. Finally, each unit cell is read out by a dedicated charge sensor that measures the parity of the three corresponding vertical pairs sequentially, allowing simultaneous readout of all unit cells. 
\textbf{b}, Measurement results of each vertical pair using the multi-qubit sequence described in \textbf{a}, with the simultaneous control of all qubits in the bottom channel of the device (pink). For comparison, we overlay the results of a similar experiment, where an individual-qubit sequence is used (blue). The presented data are vertically offset by 1 per qubit. 
\textbf{c}, Measured spin flip probabilities $P_\text{flip}$ for all vertical pairs as a function of the addressed qubit. Each column corresponds to a measurement of the full array after a single X-gate is applied to the targeted qubit.}
\label{fig:4}
\end{figure*}

\section{Multi-qubit initialization and readout}
Operating a quantum processing unit requires the ability to initialize, control, and read out all qubits within the processor in a scalable manner. Here, we demonstrate simultaneous state preparation and readout of the full 18-qubit array using the multi-qubit sequence illustrated in Fig.~\ref{fig:4}a. Initialization and readout are organized on a unit-cell basis, with all unit cells being prepared and measured simultaneously following the same protocol. Within each unit cell, the three vertical qubit pairs are initialized sequentially: first the left pair, then the right pair, and finally the center pair. The same ordering is used during readout. Such sequential ordering avoids simultaneous initialization or readout of neighboring vertical pairs, which could otherwise introduce unwanted interactions between horizontally adjacent qubits and degrade state-preparation or readout fidelities. Also, the long spin relaxation times of the qubits ensure that their states remain preserved while other pairs in the unit cell are sequentially prepared or measured. Each charge sensor is primarily sensitive to the quantum dots within its own unit cell, which enables parallel readout across different unit cells without introducing measurement cross-talk. The protocol is executed in parallel at the unit-cell level, enabling efficient operation of the array and ensuring that SPAM overhead does not increase with the size of the quantum processor. 

To verify simultaneous initialization, control, and readout across the array, we apply microwave drives of varying duration to one qubit per vertical pair within a single sequence (see circuit in Fig.~\ref{fig:4}a). The resulting Rabi oscillations are shown in Fig.~\ref{fig:4}b, where all nine even-numbered qubits are operated simultaneously. The high visibility of the oscillations confirms successful parallel qubit initialization, control, and readout. The Rabi visibility, defined by the oscillation amplitude, is comparable between individual and simultaneous operation (see \cref{fig:vis_compare}). The minor reduction in visibility is attributed to a slight broadening of the charge-sensor Coulomb peaks induced by the additional pulsing and can be mitigated by an improved charge-sensor design.

To further examine the crosstalk, we start from the fully initialized 18-qubit state and apply an X-gate to a single qubit, followed by a parity measurement of all vertical qubit pairs using the protocol described above. This yields a parity bit string for the entire array for every measurement shot, allowing us to extract the parity probability for every qubit pair. This experiment is repeated while addressing a different qubit for every cycle. The resulting data are shown in Fig.~\ref{fig:4}c and exhibit a characteristic diagonal pattern, confirming low crosstalk in the single-qubit control, simultaneous initialization, and readout across the full array. We note that there is a small crosstalk in the charge readout of Q11,Q12 when Q5,Q6 are read out. This can be attributed to a finite sensitivity of S$_2$ to the charge state of Q5,Q6, combined with a limited sensitivity to Q11,Q12. Such charge crosstalk can be suppressed by improving the sensor response to the target qubit pair (Q11,Q12 in this case), allowing for better state thresholding. Furthermore, we note that this protocol can be extended from a parity readout on 9 qubit pairs to a full readout of all 18 qubits by using CNOT gate projections before performing the parity readout~\cite{philips2022universal}.

\begin{figure*}[htbp] 
\center{\includegraphics[width=\linewidth]{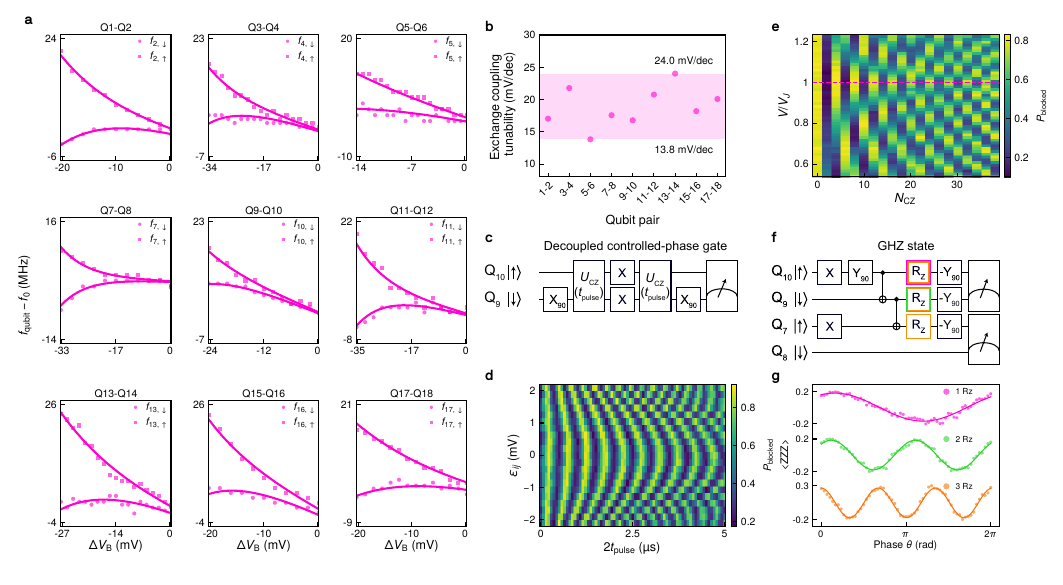}}
\caption{\textbf{Two-qubit couplings and quantum gate circuits in a 2$\times$2 plaquette.}
\textbf{a}, Two-qubit spectroscopy for all nine vertical qubit pairs in the 18-qubit array. The measurements show the splitting of the qubit resonance frequency ($f_\text{qubit}-f_0$) as a function of the applied barrier gate pulse amplitude ($\Delta V_\text{B}$), for a spin-down (circles) and spin-up (squares) initialization of the control qubit. This frequency splitting directly reflects the exchange interaction between the two qubits.
\textbf{b}, Exchange tunability for each vertical qubit pair, defined as the barrier pulse amplitude required to change the exchange coupling by one order of magnitude. The average exchange tunability is $\sim19\pm3~\text{mV/dec}$. \textbf{c}, Quantum circuit of a decoupled controlled-phase (CZ) gate interleaved within a Ramsey sequence on Q9. \textbf{d}, Return probability $P_{\rm blocked}$ after the decoupled CZ circuit, as a function of detuning $\varepsilon_{9,10}$ and total barrier pulse duration $2t_\text{pulse}$.
\textbf{e}, Amplitude calibration of the CZ gate. Return probability $P_{\rm blocked}$ by repeating $N_\text{CZ}$ times the CZ pulses in a decoupling sequence, as a function of the relative pulse amplitude $V/V_J$ and $N_\text{CZ}$. The magenta dashed line shows the calibrated amplitude of the CZ gate.
\textbf{f}, Quantum circuit for preparing a three-qubit GHZ state and measuring parity oscillations. For each experimental implementation (magenta, green, orange), only the correspondingly colored R$_z$ gates are applied.
\textbf{g}, Parity oscillations of the three-qubit GHZ state as a function of the applied phase $\theta$, for single-qubit (1 R$_z$), two-qubit (2 R$_z$) and three-qubit (3 R$_z$) phase rotations.}
\label{fig:3}
\end{figure*}

\section{Two-qubit gate and quantum algorithm}
In addition to scalable initialization and readout, universal quantum computing requires an entangling two-qubit gate. In spin qubit systems, such gates are typically implemented via the exchange interaction between nearest-neighbor qubits~\cite{loss1998quantum}, which depends exponentially on the interdot barrier potential. We characterize the exchange coupling $J$ for all vertical qubit pairs as a function of the applied barrier gate voltage $\Delta V_\text{B}$ by performing qubit spectroscopy for both basis states of the neighboring qubit. The resulting state-dependent splitting of the resonance frequency, shown in Fig.~\ref{fig:3}a, provides a direct measure of the exchange interaction. From these data, we extract an average exchange tunability of $19\pm2.9$~mV/dec (Fig.~\ref{fig:3}b), comparable to state-of-the-art results and sufficient for high-fidelity one- and two-qubit gate operations~\cite{heinz2024analysis}. 

Using this gate-controllable exchange interaction, we implement a controlled-phase (CZ) gate between qubits Q9 and Q10. The CZ gate is realized using an adiabatic, Hamming-window-shaped voltage pulse applied to the virtual interdot barrier gate~\cite{xue2022quantum,wang2024operating}. To suppress sensitivity to detuning voltage noise, the operation is performed at the zero detuning point, which is identified by embedding a decoupled CZ (DCZ) pulse within a Ramsey sequence while sweeping the detuning and pulse duration. The corresponding quantum circuit is shown in Fig.~\ref{fig:3}c and the resulting fingerprint oscillations are presented in Fig.~\ref{fig:3}d. At zero detuning ($\varepsilon_{9,10}=0$), we observe high-quality exchange oscillations, indicative of a high-fidelity two-qubit gate.

We realize a CZ gate between Q9 and Q10 by calibrating both the single qubit phases (due to $g$-tensor modulation) and the conditional phase acquired during the interaction. Fig.~\ref{fig:3}e shows the conditional phase calibration by applying repeated CZ operations, where the absence of beating at $V/V_J=1$ confirms proper phase alignment and correct gate calibration. The same procedure is applied to the horizontal qubit pair Q7,Q9 (see \cref{fig:2by2coup}), demonstrating two-qubit control along both directions of the array. To demonstrate the connectivity in this two-dimensional architecture, we generate a three-qubit Greenberger-Horne-Zeilinger (GHZ) state on qubits Q7, Q9, and Q10 by applying the circuit shown in Fig.~\ref{fig:3}f. The required CNOT operations are decomposed into combinations of native Y and CZ gates~\cite{philips2022universal}. We verify the generation of the GHZ state by measuring parity oscillations as a function of applied single-qubit phases. The effective ZZZ measurement is implemented using two parity readouts on qubit pairs Q7,Q8 and Q9,Q10, with Q8 acting as an ancilla (see Methods). The resulting oscillations are shown in Fig.~\ref{fig:3}g, where the oscillation frequency scales with the number of qubits on which the phase is varied, confirming the presence of multi-qubit entanglement.

\section{Conclusion}
In summary, we have demonstrated the operation of a modular 18-qubit array based on an extendable architecture of germanium spin qubits, featuring simultaneous initialization, control and readout across the full array. The architecture enables parallel operation on a unit-cell basis, such that the overhead associated with state preparation and qubit measurement does not increase with system size. We achieve high-fidelity single-qubit control across all qubits, tunable exchange coupling, and implement two-qubit gates that enable the generation of multi-qubit entangled states across both directions of the two-dimensional array. These results establish a scalable platform for two-dimensional semiconductor quantum processors. By leveraging advanced semiconductor manufacturing to scale the number of qubits in both dimensions and improve device uniformity, together with the integration of cryogenic control electronics, this architecture provides a viable pathway toward fault-tolerant quantum computation based on semiconductor spin qubits.

\section*{Acknowledgements}
We thank all team members from the spin-qubit groups at QuTech for the helpful discussions, collective setup debugging, and general interactions, particularly the in-depth discussions with Irene Fernandez de Fuentes, Yuta Matsumoto, and Maximilian Rimbach-Russ. Furthermore, we are grateful for software support and useful discussions with Sander de Snoo. We also thank QuTech technical staff for their support in building and maintaining our experimental setups. Finally, we thank the staff of the Kavli Nanolab Delft for cleanroom support. We acknowledge the support of the European Union through the EIC Transition Grant GROOVE 101113173. The authors also acknowledge support from Holland High Tech through the PPS TKI-HTSM project HiLoGe. This publication is supported by QuTech NWO funding Part III 'Application based research – Demonstrators', with Project number 601.QT.001 Part III-C\_Groove, financed by the Dutch Research Council (NWO).

\section*{Author contributions}
J.J.D., X.Z., and N.W.H. designed the experiments. J.J.D., X.Z., D.v.D., and D.G. performed the data analysis, interpretation, and presentation. A.B. fabricated the device on material provided by L.E.A.S. and G.S. D.G. performed device screening and characterization. D.B., A.B., and D.v.D. contributed to the experimental hardware setup; A.C. contributed to the experimental software setup. J.J.D., X.Z., and N.W.H. wrote the manuscript with input from all co-authors. A.-M.J.Z. and N.W.H. supervised the project.

\section*{Competing interests}
N.W.H. and A.-M.J.Z are directors of Groove Quantum B.V. A.B., A.C., A-M.J.Z., D.B., D.v.D., D.G., G.S., J.D., N.W.H., and X.Z. declare equity interest in Groove Quantum. The other authors declare no competing interests.

\section*{Methods}
\label{sec:methods}
\subsection*{Device and setup}
The quantum processor is fabricated on a Ge/SiGe heterostructure with a buried strained germanium quantum well. The device contains an overlapping gate stack, consisting of three separate electrically isolated gate layers, and ohmic contacts to the quantum well. The device is thermally anchored to the mixing chamber of a dilution refrigerator with a base temperature of around 14 mK. The charge sensors are measured using radio-frequency reflectometry, with tank circuits formed by NbTiN inductors mounted on the PCB and the spurious capacitance of the bonding wires and metal lines on the board and chip. Sensor reflectometry is performed using Qblox QRM modules. All plunger and barrier gates of the device are connected to Qblox QCM modules that generate all control signals.

\subsection*{Initialization protocol}
To intialize the full QPU, we follow the conceptual protocol described in the main text. For each readout pair, we first align the DQD detuning to a relaxation hot spot to allow for a fast decay into the lowest-energy even parity charge state (e.g. S(2,0) or S(0,2)). Next, we take a reference measurement to account for slow charge sensor drifts and adiabatically ramp the DQD detuning to the middle of the odd parity charge state (e.g. (1,1)), thereby initializing either a $\ket{\downarrow\downarrow}$ or $\ket{\downarrow\uparrow}$ spin state, depending on the relative size of the different anti-crossings~\cite{tsoukalas2026dressed}. The initialization state for the entire device is $\ket{\downarrow\downarrow\downarrow\downarrow\downarrow\downarrow\uparrow\downarrow\downarrow\uparrow\downarrow\uparrow\downarrow\downarrow\downarrow\downarrow\uparrow\downarrow}$, which is inferred from the sign of the frequency shifts observed in the two-qubit spectroscopy (Fig.~\ref{fig:3}a).

\subsection*{Randomized benchmarking}
We perform randomized benchmarking to measure the single-qubit fidelities of the 18 qubits reported in Fig.~\ref{fig:1}b. For each qubit, the data are extracted as the difference between the benchmarking measurements with target states  $\ket{0}$ and $\ket{1}$. The resulting randomized benchmarking data and fits are shown in \cref{fig:RBs_diff}. To fit the data points, we assume that they follow an exponential decay of the form $P = AF^n + B$, with $A$ and $B$ as fitting parameters, $F$ as the circuit level fidelity, and $n$ as the number of Clifford operations. To calculate the Clifford fidelity, we use the following expression: 
\begin{equation}
F_C = 1-(1-F)/2
\label{eq:F_clifford}
\end{equation}
while the native gate fidelity is defined by:
\begin{equation}
F_\text{gate} = 1-(1-F_C)/(2N_\text{avg})
\label{eq:F_gate}
\end{equation}
in which $N_\text{avg}$ is the average gate number per Clifford and is equal to 2 for our chosen Cliffords~\cite{john2024two}. Importantly, we measured the fidelities using the same voltage settings across the entire 18-qubit array.

\subsection*{Gate virtualization}
We employ a four-layer gate-virtualization scheme to mitigate control crosstalk~\cite{rao2025modular}. In the first layer, we compensate the the influence of the device gates on the charge sensors. In the second layer, we orthogonalize the plunger gates, enabling independent control of the chemical potential of each quantum dot. In the third layer, we normalize the charging energies of the quantum dots. Finally, in the fourth layer, we compensate the cross talk from the interdot barrier gates onto the quantum dot chemical potentials.

\subsection*{Parity readout}
In the GHZ-state measurements, the parity readouts of (Q7,Q8) and (Q9,Q10) provide access to a four-body correlator. From the joint parity outcomes, we extract the expectation value $\langle Z_7Z_8Z_9Z_{10} \rangle=P_\text{EE}-P_\text{EO}-P_\text{OE}+P_\text{OO}$, where $P_\text{EE}$, $P_\text{EO}$, $P_\text{OE}$ and $P_\text{OO}$ are the joint probabilities for the even/even, even/odd, odd/even and odd/odd parity outcomes, respectively. Since Q8 is initialized in $\ket{\downarrow}$ and remains unchanged throughout the circuit, $\langle Z_8 \rangle=-1$, such that $\langle Z_7Z_9Z_{10} \rangle=-\langle Z_7Z_8Z_9Z_{10} \rangle$.

\subsection*{Automatic calibration}
To compensate for slow drifts and ensure stable device operation, an automated calibration procedure is executed regularly. This procedure is structured as a sequence of deterministic calibration steps, each followed by automated evaluation of the data and updating of the corresponding calibration parameters. \cref{fig:auto_cali} illustrates the calibration flow applied to our 18-qubit device, highlighting the sequence of routines required to track and correct daily fluctuations in the system. The procedure starts with a sensor-to-flank calibration, where all charge sensors are tuned at their most sensitive operating points. This is followed by a series of Rabi-based calibrations on individual qubits, including optimization of the readout sequence, qubit drive frequency, qubit drive amplitude, and state discrimination thresholds.

As each calibration stage is dependent on the successful completion of the preceding one, we implement a structured and hierarchical workflow. The calibration diagram, automatically generated as part of the calibration routine, provides a compact visual summary of the calibration status, enabling a quick verification of the successful execution of all steps. This methodology supports a gradual transition from manual to fully automated calibration while retaining the possibility for human oversight and intervention when needed.
\bibliography{library}

@PREAMBLE{
 "\providecommand{\noopsort}[1]{}" 
 # "\providecommand{\singleletter}[1]{#1}%" 
}

@article{hendrickx2021four,
  title={A four-qubit germanium quantum processor},
  author={Hendrickx, Nico W and Lawrie, William IL and Russ, Maximilian and van Riggelen, Floor and de Snoo, Sander L and Schouten, Raymond N and Sammak, Amir and Scappucci, Giordano and Veldhorst, Menno},
  journal={Nature},
  volume={591},
  number={7851},
  pages={580--585},
  year={2021},
  publisher={Nature Publishing Group UK London}
}

@article{philips2022universal,
  title={Universal control of a six-qubit quantum processor in silicon},
  author={Philips, Stephan GJ and M{\k{a}}dzik, Mateusz T and Amitonov, Sergey V and de Snoo, Sander L and Russ, Maximilian and Kalhor, Nima and Volk, Christian and Lawrie, William IL and Brousse, Delphine and Tryputen, Larysa and others},
  journal={Nature},
  volume={609},
  number={7929},
  pages={919--924},
  year={2022},
  publisher={Nature Publishing Group UK London}
}

@article{john2024two,
  title={Robust and localised control of a 10-spin qubit array in germanium},
  author={John, Valentin and Yu, C{\'e}cile X and van Straaten, Barnaby and Rodr{\'\i}guez-Mena, Esteban A and Rodr{\'\i}guez, Mauricio and Oosterhout, Stefan D and Stehouwer, Lucas EA and Scappucci, Giordano and Rimbach-Russ, Maximilian and Bosco, Stefano and others},
  journal={Nature Communications},
  volume={16},
  number={1},
  pages={10560},
  year={2025},
  publisher={Nature Publishing Group UK London}
}

@article{wu2025simultaneous,
  title={Simultaneous High-Fidelity Single-Qubit Gates in a Spin Qubit Array},
  author={Wu, Yi-Hsien and Camenzind, Leon C and B{\"u}tler, Patrick and Jin, Ik Kyeong and Noiri, Akito and Takeda, Kenta and Nakajima, Takashi and Kobayashi, Takashi and Scappucci, Giordano and Goan, Hsi-Sheng and others},
  journal={arXiv preprint arXiv:2507.11918},
  year={2025}
}

@article{xue2022quantum,
  title={Quantum logic with spin qubits crossing the surface code threshold},
  author={Xue, Xiao and Russ, Maximilian and Samkharadze, Nodar and Undseth, Brennan and Sammak, Amir and Scappucci, Giordano and Vandersypen, Lieven MK},
  journal={Nature},
  volume={601},
  number={7893},
  pages={343--347},
  year={2022},
  publisher={Nature Publishing Group UK London}
}

@article{scappucci2021germanium,
  title={The germanium quantum information route},
  author={Scappucci, Giordano and Kloeffel, Christoph and Zwanenburg, Floris A and Loss, Daniel and Myronov, Maksym and Zhang, Jian-Jun and De Franceschi, Silvano and Katsaros, Georgios and Veldhorst, Menno},
  journal={Nature Reviews Materials},
  volume={6},
  number={10},
  pages={926--943},
  year={2021},
  publisher={Nature Publishing Group UK London}
}

@article{neyens2024probing,
  title={Probing single electrons across 300-mm spin qubit wafers},
  author={Neyens, Samuel and Zietz, Otto K and Watson, Thomas F and Luthi, Florian and Nethwewala, Aditi and George, Hubert C and Henry, Eric and Islam, Mohammad and Wagner, Andrew J and Borjans, Felix and others},
  journal={Nature},
  volume={629},
  number={8010},
  pages={80--85},
  year={2024},
  publisher={Nature Publishing Group UK London}
}

@article{zwerver2022qubits,
  title={Qubits made by advanced semiconductor manufacturing},
  author={Zwerver, AMJ and Kr{\"a}henmann, T and Watson, TF and Lampert, Lester and George, Hubert C and Pillarisetty, Ravi and Bojarski, SA and Amin, Payam and Amitonov, SV and Boter, JM and others},
  journal={Nature Electronics},
  volume={5},
  number={3},
  pages={184--190},
  year={2022},
  publisher={Nature Publishing Group UK London}
}

@article{stano2022review,
  title={Review of performance metrics of spin qubits in gated semiconducting nanostructures},
  author={Stano, Peter and Loss, Daniel},
  journal={Nature Reviews Physics},
  volume={4},
  number={10},
  pages={672--688},
  year={2022},
  publisher={Nature Publishing Group UK London}
}

@article{george2025intel12qb,
  title={12-spin-qubit arrays fabricated on a 300 mm semiconductor manufacturing line},
  author={George, Hubert C and Madzik, Mateusz T and Henry, Eric M and Wagner, Andrew J and Islam, Mohammad M and Borjans, Felix and Connors, Elliot J and Corrigan, Joelle and Curry, Matthew and Harper, Michael K and others},
  journal={Nano Letters},
  volume={25},
  number={2},
  pages={793--799},
  year={2024},
  publisher={ACS Publications}
}

@article{mohseni2024build,
  title={How to build a quantum supercomputer: Scaling from hundreds to millions of qubits},
  author={Mohseni, Masoud and Scherer, Artur and Johnson, K Grace and Wertheim, Oded and Otten, Matthew and Aadit, Navid Anjum and Alexeev, Yuri and Bresniker, Kirk M and Camsari, Kerem Y and Chapman, Barbara and others},
  journal={arXiv preprint arXiv:2411.10406},
  year={2024}
}

@article{steinacker2025industry,
  title={Industry-compatible silicon spin-qubit unit cells exceeding 99\% fidelity},
  author={Steinacker, Paul and Dumoulin Stuyck, Nard and Lim, Wee Han and Tanttu, Tuomo and Feng, MengKe and Serrano, Santiago and Nickl, Andreas and Candido, Marco and Cifuentes, Jesus D and Vahapoglu, Ensar and others},
  journal={Nature},
  pages={1--7},
  year={2025},
  publisher={Nature Publishing Group UK London}
}

@article{xue2021cmos,
  title={CMOS-based cryogenic control of silicon quantum circuits},
  author={Xue, Xiao and Patra, Bishnu and van Dijk, Jeroen PG and Samkharadze, Nodar and Subramanian, Sushil and Corna, Andrea and Paquelet Wuetz, Brian and Jeon, Charles and Sheikh, Farhana and Juarez-Hernandez, Esdras and others},
  journal={Nature},
  volume={593},
  number={7858},
  pages={205--210},
  year={2021},
  publisher={Nature Publishing Group UK London}
}

@article{bartee2025spin,
  title={Spin-qubit control with a milli-kelvin CMOS chip},
  author={Bartee, Samuel K and Gilbert, Will and Zuo, Kun and Das, Kushal and Tanttu, Tuomo and Yang, Chih Hwan and Dumoulin Stuyck, Nard and Pauka, Sebastian J and Su, Rocky Y and Lim, Wee Han and others},
  journal={Nature},
  pages={1--6},
  year={2025},
  publisher={Nature Publishing Group UK London}
}

@article{huang2024high,
  title={High-fidelity spin qubit operation and algorithmic initialization above 1 K},
  author={Huang, Jonathan Y and Su, Rocky Y and Lim, Wee Han and Feng, MengKe and Van Straaten, Barnaby and Severin, Brandon and Gilbert, Will and Dumoulin Stuyck, Nard and Tanttu, Tuomo and Serrano, Santiago and others},
  journal={Nature},
  volume={627},
  number={8005},
  pages={772--777},
  year={2024},
  publisher={Nature Publishing Group UK London}
}

@article{zhang2025universal,
  title={Universal control of four singlet--triplet qubits},
  author={Zhang, Xin and Morozova, Elizaveta and Rimbach-Russ, Maximilian and Jirovec, Daniel and Hsiao, Tzu-Kan and Fari{\~n}a, Pablo Cova and Wang, Chien-An and Oosterhout, Stefan D and Sammak, Amir and Scappucci, Giordano and others},
  journal={Nature Nanotechnology},
  volume={20},
  number={2},
  pages={209--215},
  year={2025},
  publisher={Nature Publishing Group UK London}
}

@article{hendrickx2024sweet,
  title={Sweet-spot operation of a germanium hole spin qubit with highly anisotropic noise sensitivity},
  author={Hendrickx, NW and Massai, Leonardo and Mergenthaler, Matthias and Schupp, Felix Julian and Paredes, Stephan and Bedell, SW and Salis, G and Fuhrer, A},
  journal={Nature Materials},
  volume={23},
  number={7},
  pages={920--927},
  year={2024},
  publisher={Nature Publishing Group UK London}
}

@article{yu2025optimising,
  title={Optimising germanium hole spin qubits with a room-temperature magnet},
  author={Yu, Cecile X and van Straaten, Barnaby and Ivlev, Alexander S and John, Valentin and Oosterhout, Stefan D and Stehouwer, Lucas EA and Borsoi, Francesco and Scappucci, Giordano and Veldhorst, Menno},
  journal={arXiv preprint arXiv:2507.03390},
  year={2025}
}

@article{rimbach2023simple,
  title={Simple framework for systematic high-fidelity gate operations},
  author={Rimbach-Russ, Maximilian and Philips, Stephan GJ and Xue, Xiao and Vandersypen, Lieven MK},
  journal={Quantum Science and Technology},
  volume={8},
  number={4},
  pages={045025},
  year={2023},
  publisher={IOP Publishing}
}

@article{stehouwer2025exploiting,
  author    = {Stehouwer, Lucas E. A. and Yu, C{\'e}cile X. and van Straaten, Barnaby and Tosato, Alberto and John, Valentin and Degli Esposti, Davide and Elsayed, Asser and Costa, Davide and Oosterhout, Stefan D. and Hendrickx, Nico W. and Veldhorst, Menno and Borsoi, Francesco and Scappucci, Giordano},
  title     = {Exploiting strained epitaxial germanium for scaling low-noise spin qubits at the micrometre scale},
  journal   = {Nature Materials},
  year      = {2025},
  volume    = {24},
  number    = {12},
  pages     = {1906--1912},
  doi       = {10.1038/s41563-025-02276-w}
}

@article{heinz2024analysis,
  title={Analysis and mitigation of residual exchange coupling in linear spin-qubit arrays},
  author={Heinz, Irina and Mills, Adam R and Petta, Jason R and Burkard, Guido},
  journal={Physical Review Research},
  volume={6},
  number={1},
  pages={013153},
  year={2024},
  publisher={APS}
}

@article{franke2019rent,
  title={Rent’s rule and extensibility in quantum computing},
  author={Franke, David P and Clarke, James S and Vandersypen, Lieven MK and Veldhorst, Menno},
  journal={Microprocessors and Microsystems},
  volume={67},
  pages={1--7},
  year={2019},
  publisher={Elsevier}
}

@article{vandersypen2017interfacing,
  title={Interfacing spin qubits in quantum dots and donors—hot, dense, and coherent},
  author={Vandersypen, Lieven MK and Bluhm, Hendrik and Clarke, JS and Dzurak, AS and Ishihara, R and Morello, A and Reilly, DJ and Schreiber, LR and Veldhorst, M},
  journal={npj Quantum Information},
  volume={3},
  number={1},
  pages={34},
  year={2017},
  publisher={Nature Publishing Group UK London}
}

@article{wang2024operating,
  title={Operating semiconductor quantum processors with hopping spins},
  author={Wang, Chien-An and John, Valentin and Tidjani, Hanifa and Yu, C{\'e}cile X and Ivlev, Alexander S and D{\'e}prez, Corentin and van Riggelen-Doelman, Floor and Woods, Benjamin D and Hendrickx, Nico W and Lawrie, William IL and others},
  journal={Science},
  volume={385},
  number={6707},
  pages={447--452},
  year={2024},
  publisher={American Association for the Advancement of Science}
}

@article{hensgens2017quantum,
  title={Quantum simulation of a Fermi--Hubbard model using a semiconductor quantum dot array},
  author={Hensgens, Toivo and Fujita, Takafumi and Janssen, Laurens and Li, Xiao and Van Diepen, CJ and Reichl, Christian and Wegscheider, Werner and Das Sarma, Sankar and Vandersypen, Lieven MK},
  journal={Nature},
  volume={548},
  number={7665},
  pages={70--73},
  year={2017},
  publisher={Nature Publishing Group UK London}
}

@article{fischer2008spin,
  title = {Spin decoherence of a heavy hole coupled to nuclear spins in a quantum dot},
  author = {Fischer, Jan and Coish, W. A. and Bulaev, D. V. and Loss, Daniel},
  journal = {Phys. Rev. B},
  volume = {78},
  issue = {15},
  pages = {155329},
  numpages = {9},
  year = {2008},
  month = {Oct},
  publisher = {American Physical Society}
}

@article{tsoukalas2026dressed,
  title={A dressed singlet-triplet qubit in germanium},
  author={Tsoukalas, Konstantinos and von L{\"u}pke, Uwe and Orekhov, Alexei and Het{\'e}nyi, Bence and Seidler, Inga and Sommer, Lisa and Kelly, Eoin G and Massai, Leonardo and Aldeghi, Michele and Pita-Vidal, Marta and others},
  journal={Nature Communications},
  volume={17},
  number={1},
  pages={699},
  year={2026},
  publisher={Nature Publishing Group UK London}
}

@article{veldhorst2017silicon,
  title={Silicon CMOS architecture for a spin-based quantum computer},
  author={Veldhorst, Menno and Eenink, Harm GJ and Yang, Chih-Hwan and Dzurak, Andrew S},
  journal={Nature communications},
  volume={8},
  number={1},
  pages={1766},
  year={2017},
  publisher={Nature Publishing Group UK London}
}

@article{undseth2026parity,
  title={Weight-four parity checks with silicon spin qubits},
  author={Undseth, Brennan and Meggiato, Nicola and Wu, Yi-Hsien and Katiraee-Far, Sam R and Tryputen, Larysa and de Snoo, Sander L and Esposti, Davide Degli and Scappucci, Giordano and Greplov{\'a}, Eli{\v{s}}ka and Vandersypen, Lieven MK},
  journal={arXiv preprint arXiv:2601.23267},
  year={2026}
}

@article{matsumoto2025two,
  title={Two-qubit logic and teleportation with mobile spin qubits in silicon},
  author={Matsumoto, Yuta and De Smet, Maxim and Tryputen, Larysa and de Snoo, Sander L and Amitonov, Sergey V and Sammak, Amir and Rimbach-Russ, Maximilian and Scappucci, Giordano and Vandersypen, Lieven MK},
  journal={arXiv preprint arXiv:2503.15434},
  year={2025}
}

@article{ademi2025distributing,
  title={Distributing entanglement between distant semiconductor qubit registers using a shared-control shuttling link},
  author={Ademi, Zarije and Bassi, Marion and Yu, C{\'e}cile X and Oosterhout, Stefan D and Matsumoto, Yuta and de Snoo, Sander L and Sammak, Amir and Vandersypen, Lieven MK and Scappucci, Giordano and D{\'e}prez, Corentin and others},
  journal={arXiv preprint arXiv:2510.26860},
  year={2025}
}

@article{nickl2025eight,
  title={Eight-Qubit Operation of a 300 mm SiMOS Foundry-Fabricated Device},
  author={Nickl, Andreas and Stuyck, Nard Dumoulin and Steinacker, Paul and Cifuentes, Jesus D and Serrano, Santiago and Feng, MengKe and Vahapoglu, Ensar and Hudson, Fay E and Chan, Kok Wai and Kubicek, Stefan and others},
  journal={arXiv preprint arXiv:2512.10174},
  year={2025}
}

@article{seidler2025spatial,
  title={Spatial uniformity of g-tensor and spin-orbit interaction in germanium hole spin qubits},
  author={Seidler, Inga and Het{\'e}nyi, Bence and Sommer, Lisa and Massai, Leonardo and Tsoukalas, Konstantinos and Kelly, Eoin G and Orekhov, Alexei and Aldeghi, Michele and Bedell, Stephen W and Paredes, Stephan and others},
  journal={arXiv preprint arXiv:2510.03125},
  year={2025}
}

@article{gonzalez2021scaling,
  title={Scaling silicon-based quantum computing using CMOS technology},
  author={Gonzalez-Zalba, MF and De Franceschi, S and Charbon, E and Meunier, Tristan and Vinet, M and Dzurak, AS},
  journal={Nature Electronics},
  volume={4},
  number={12},
  pages={872--884},
  year={2021},
  publisher={Nature Publishing Group UK London}
}

@article{loss1998quantum,
  title={Quantum computation with quantum dots},
  author={Loss, Daniel and DiVincenzo, David P},
  journal={Physical Review A},
  volume={57},
  number={1},
  pages={120},
  year={1998},
  publisher={APS}
}

@article{takeda2022quantum,
  title={Quantum error correction with silicon spin qubits},
  author={Takeda, Kenta and Noiri, Akito and Nakajima, Takashi and Kobayashi, Takashi and Tarucha, Seigo},
  journal={Nature},
  volume={608},
  number={7924},
  pages={682--686},
  year={2022},
  publisher={Nature Publishing Group UK London}
}

@article{zhang2026quantum,
  title={Quantum error detection in a silicon quantum processor},
  author={Zhang, Chunhui and Li, Chunhui and Tian, Zhen and Jiang, Yan and Xu, Feng and Zhang, Shihang and Wang, Hao and Zhang, Yu-Ning and Bai, Xuesong and Zhao, Baolong and others},
  journal={Nature Electronics},
  pages={1--9},
  year={2026},
  publisher={Nature Publishing Group UK London}
}

@article{lodari2021low,
  title={Low percolation density and charge noise with holes in germanium},
  author={Lodari, Mario and Hendrickx, Nico W and Lawrie, William IL and Hsiao, Tzu-Kan and Vandersypen, Lieven MK and Sammak, Amir and Veldhorst, Menno and Scappucci, Giordano},
  journal={Materials for Quantum Technology},
  volume={1},
  number={1},
  pages={011002},
  year={2021},
  publisher={IOP Publishing}
}

@article{mills2022two,
  title={Two-qubit silicon quantum processor with operation fidelity exceeding 99\%},
  author={Mills, Adam R and Guinn, Charles R and Gullans, Michael J and Sigillito, Anthony J and Feldman, Mayer M and Nielsen, Erik and Petta, Jason R},
  journal={Science Advances},
  volume={8},
  number={14},
  pages={eabn5130},
  year={2022},
  publisher={American Association for the Advancement of Science}
}

@article{van2022phase,
  title={Phase flip code with semiconductor spin qubits},
  author={Van Riggelen, F and Lawrie, WIL and Russ, M and Hendrickx, NW and Sammak, A and Rispler, M and Terhal, BM and Scappucci, G and Veldhorst, M},
  journal={npj Quantum information},
  volume={8},
  number={1},
  pages={124},
  year={2022},
  publisher={Nature Publishing Group UK London}
}

@article{de2025high,
  title={High-fidelity single-spin shuttling in silicon},
  author={De Smet, Maxim and Matsumoto, Yuta and Zwerver, Anne-Marije J and Tryputen, Larysa and de Snoo, Sander L and Amitonov, Sergey V and Katiraee-Far, Sam R and Sammak, Amir and Samkharadze, Nodar and G{\"u}l, {\"O}nder and others},
  journal={Nature Nanotechnology},
  volume={20},
  number={7},
  pages={866--872},
  year={2025},
  publisher={Nature Publishing Group UK London}
}

@article{rao2025modular,
  title={Modular autonomous virtualization system for two-dimensional semiconductor quantum dot arrays},
  author={Rao, Anantha S and Buterakos, Donovan and van Straaten, Barnaby and John, Valentin and Yu, C{\'e}cile X and Oosterhout, Stefan D and Stehouwer, Lucas and Scappucci, Giordano and Veldhorst, Menno and Borsoi, Francesco and others},
  journal={Physical Review X},
  volume={15},
  number={2},
  pages={021034},
  year={2025},
  publisher={APS}
}

@article{zhang2026universal,
  title={Universal logical operations in a silicon quantum processor},
  author={Zhang, Chunhui and Xu, Feng and Zhang, Shihang and Duan, Mingchao and Zhong, Dupeng and Bai, Xuesong and Wang, Hao and Huang, Chao and Deng, Yi and Gao, Miao and others},
  journal={Nature Nanotechnology},
  pages={1--7},
  year={2026},
  publisher={Nature Publishing Group UK London}
}

@article{edlbauer202511,
  title={An 11-qubit atom processor in silicon},
  author={Edlbauer, Hermann and Wang, Junliang and Huq, AM Saffat-Ee and Thorvaldson, Ian and Jones, Michael T and Misha, Saiful Haque and Pappas, William J and Moehle, Christian M and Hsueh, Yu-Ling and Bornemann, Henric and others},
  journal={Nature},
  volume={648},
  number={8094},
  pages={569--575},
  year={2025},
  publisher={Nature Publishing Group UK London}
}

@article{mkadzik2022precision,
  title={Precision tomography of a three-qubit donor quantum processor in silicon},
  author={M{\k{a}}dzik, Mateusz T and Asaad, Serwan and Youssry, Akram and Joecker, Benjamin and Rudinger, Kenneth M and Nielsen, Erik and Young, Kevin C and Proctor, Timothy J and Baczewski, Andrew D and Laucht, Arne and others},
  journal={Nature},
  volume={601},
  number={7893},
  pages={348--353},
  year={2022},
  publisher={Nature Publishing Group UK London}
}

\appendix
\makeatletter
\setcounter{section}{0}
\setcounter{figure}{0}
\setcounter{table}{0}
\setcounter{equation}{0}

\renewcommand\thefigure{\arabic{figure}}
\renewcommand\thetable{S\arabic{table}}
\renewcommand\theequation{S\arabic{equation}}

\def\appendixname{}
\def\appendixpagename{}
\def\appendixtocname{}
\@removefromreset{equation}{section}
\makeatother

\onecolumngrid

\clearpage
\makeatletter
\def\fnum@figure{\textbf{Extended Data Fig.~\thefigure}}
\makeatother

\section*{EXTENDED DATA}
\label{sec:bilinear_array}
\renewcommand{\thefigure}{\arabic{figure}}
\setcounter{figure}{0}

\begin{figure*}[htbp] 
\centering
\includegraphics[width=\linewidth]{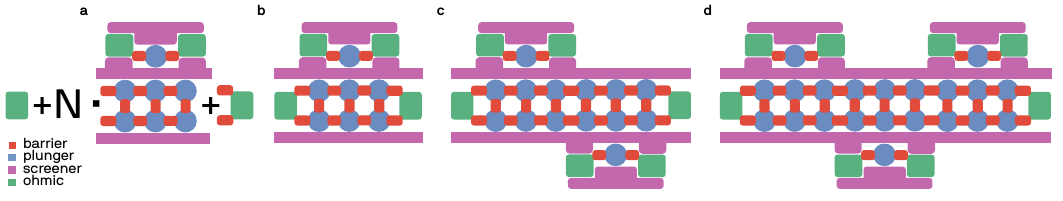}
\caption{\textbf{Extendable $\bm{2\times N}$ quantum dot array in germanium.}
\textbf{a}, Schematic of the generalized $2\times N$ quantum dot architecture. The 18-qubit array studied in the main text, is based on a repeating 2x3 unit cell that can be extended along one dimension. Each unit cell includes a dedicated charge sensor, while ohmic contacts on either end of the array act as reservoirs to load holes into the system. This modular design enables the realization of arrays with various sizes (e.g. 6, 12, or 18 qubits) and can be extended into larger systems.
\textbf{b-d}, Examples of layouts generated from the 2xN architecture: \textbf{b}, 2x3 array; \textbf{c}, 2x6 array; \textbf{d}, 2x9 array, corresponding to the device studied in the main text (see Fig.~\ref{fig:1}a). Ohmic contacts, screening gates, plunger gates, and barrier gates are indicated in green, magenta, blue and red, respectively.}
\label{fig:2xNscheme}
\end{figure*}

\begin{figure*}[htbp] 
\centering
\includegraphics[width=\linewidth]{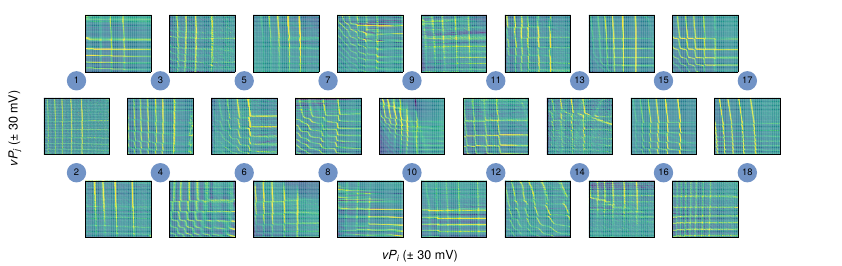}
\caption{\textbf{Horizontal and vertical double quantum dot charge stability diagrams in the 18-qubit array.} Charge stability diagrams of all double quantum dot (DQD) pairs, measured as a function of the virtual plunger gates voltages of two neighboring dots ($vP_i$, $vP_j$), for both horizontal (e.g. Q1,Q3) and vertical (e.g. Q1,Q2) pairs. Blue circles indicate the plungers gate defining the quantum dots. The gate virtualization minimizes crosstalk between control gates, ensuring that tuning a given dot does not significantly shift the chemical potentials of neighboring dots. For all DQD pairs, the top-right corner in the CSD, corresponds to the (0,0) charge configuration.}
\label{fig:all_DQD_plots}
\end{figure*}

\begin{figure*}[htbp] 
\center{\includegraphics[width=\linewidth]{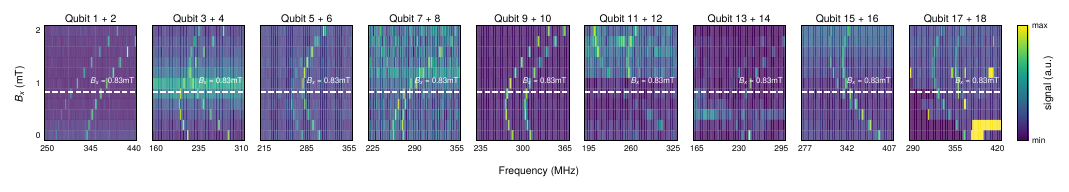}}
\caption{\textbf{Magnetic field sweep and identification of the operating conditions for qubit control.} The out-of-plane magnetic field $B_x$ is swept between 0 and 2 mT to identify optimal operating conditions, as defined by the sweet plane of minimal hyperfine interaction at the $g$-tensor equation~\cite{hendrickx2024sweet}. Due to the finite variations in qubit $g$-tensors, a single global sweet spot in the $B_x$-field cannot be found for all qubits. A field of $B_x$ = 0.83 mT (white dashed lines) is therefore chosen as a compromise, placing most qubits close to their respective sweet spots.}
\label{fig:Bx_field_sweep}
\end{figure*}

\begin{figure*}[htbp] 
\center{\includegraphics[width=\linewidth]{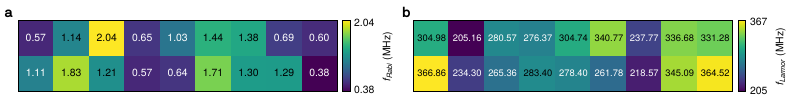}}
\caption{\textbf{Rabi and Larmor frequencies in the 18-qubit array.} Colour maps of the extracted Rabi frequencies (\textbf{a}) and Larmor frequencies (\textbf{b}), obtained from the measurements shown in Fig.~\ref{fig:2}a.}
\label{fig:18QB_Rabi&Larmor_frequencies}
\end{figure*}

\begin{figure*}[htbp] 
\center{\includegraphics[width=\linewidth]{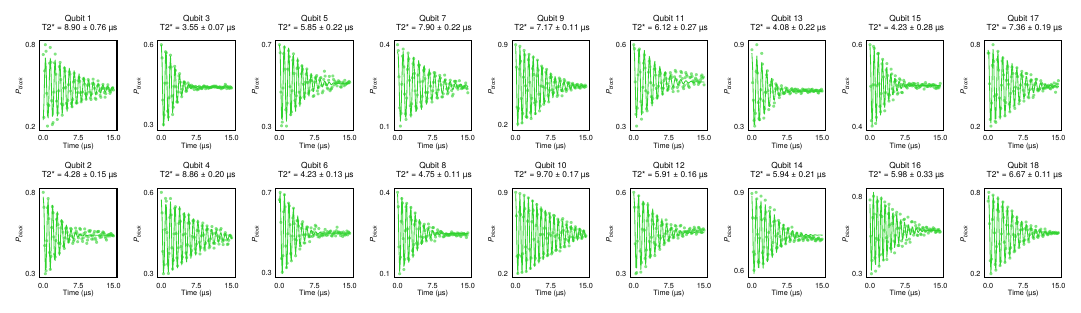}}
\caption{\textbf{Extracted $T_2^*$ values in the 18-qubit array.} Ramsey measurements used to extract qubit coherence across the array. For each qubit, the corresponding dephasing time $T_2^*$ is extracted by fitting the data to $
P_\text{block} = Ae^{-(t/T_2^*)^2}\cos{(2 \pi f t + \phi)} + C$
where $A$, $\phi$, $f$ and $C$ are fit parameters.}
\label{fig:T2star}
\end{figure*}

\begin{figure*}[htbp] 
\center{\includegraphics[width=\linewidth]{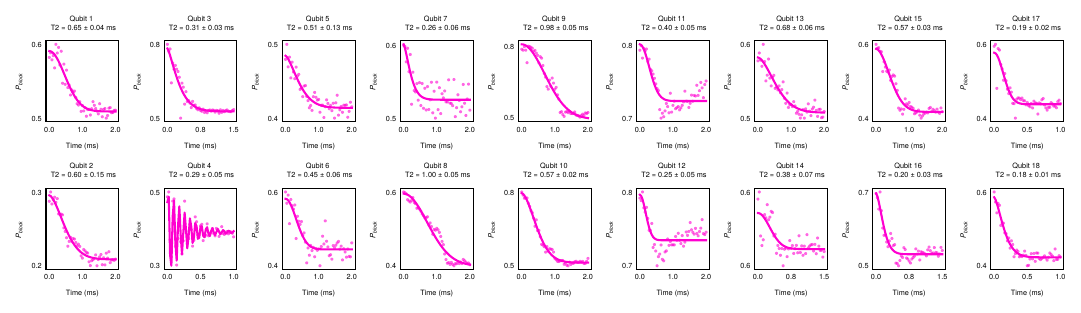}}
\caption{\textbf{Extracted $T_2^\text{CPMG}$ values in the 18-qubit array.} CPMG measurements used to extract qubit coherence across the array, using a Carr-Purcell-Meiboom-Gill sequence with $N=256$ refocussing pulses. For each qubit, the corresponding decoherence time $T_2^\text{CPMG}$ is extracted by fitting the data to:
$P_\text{block} = Ae^{-(t/T_2^\text{CPMG})^{\alpha}} + C$
where $A$, $\alpha$, and $C$ are fit parameters. We note that qubit Q4 suffers from coupling to a coherent noise source, resulting in an additional oscillation in the decay. To allow us to extract the envelope decay, we include an oscillatory component in the fit function.}
\label{fig:T2CPMG}
\end{figure*}

\begin{figure*}[htbp] 
\center{\includegraphics[width=\linewidth]{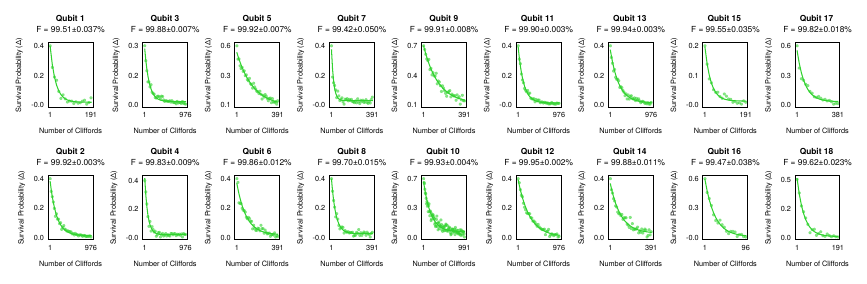}}
\caption{\textbf{Single-qubit gate randomized benchmarking in the 18-qubit array.} For every qubit, single-qubit gate randomized benchmarking results are reported. The data are extracted as the difference between randomized benchmarking measurements with target states $\ket{0}$ and $\ket{1}$ (see Methods for details).}
\label{fig:RBs_diff}
\end{figure*}

\begin{figure*}[htbp] 
\center{\includegraphics[width=0.4\linewidth]{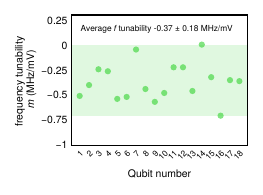}}
\caption{\textbf{Tunability of the qubit Larmor frequency, as extracted from the two-qubit gate spectroscopy.} Values of the qubit Larmor frequency tunability as a function of the vertical interdot barrier voltage, expressed as the slope of the linear component observed in the exchange spectroscopy (Fig.\ref{fig:3}a).}
\label{fig:exchange_slope}
\end{figure*}

\begin{figure*}[htbp] 
\center{\includegraphics[width=\linewidth]{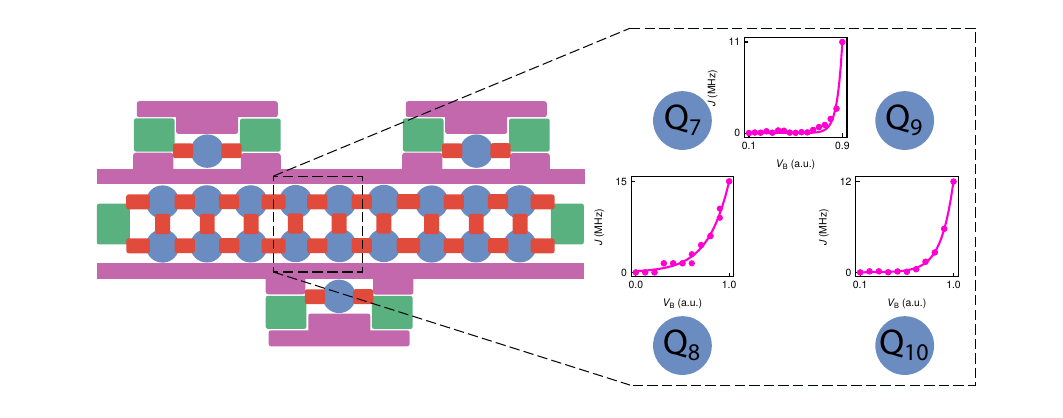}}
\caption{\textbf{Exchange coupling in a 2$\times$2 plaquette.} Illustration of gate tunability of the exchange coupling as a function of the barrier gate voltage pulse depth, for the three qubit pairs involved in the measurement of GHZ parity oscillations (Q7,Q8, Q7,Q9, and Q9,Q10).}
\label{fig:2by2coup}
\end{figure*}

\begin{figure*}[htbp] 
\center{\includegraphics[width=1\linewidth]{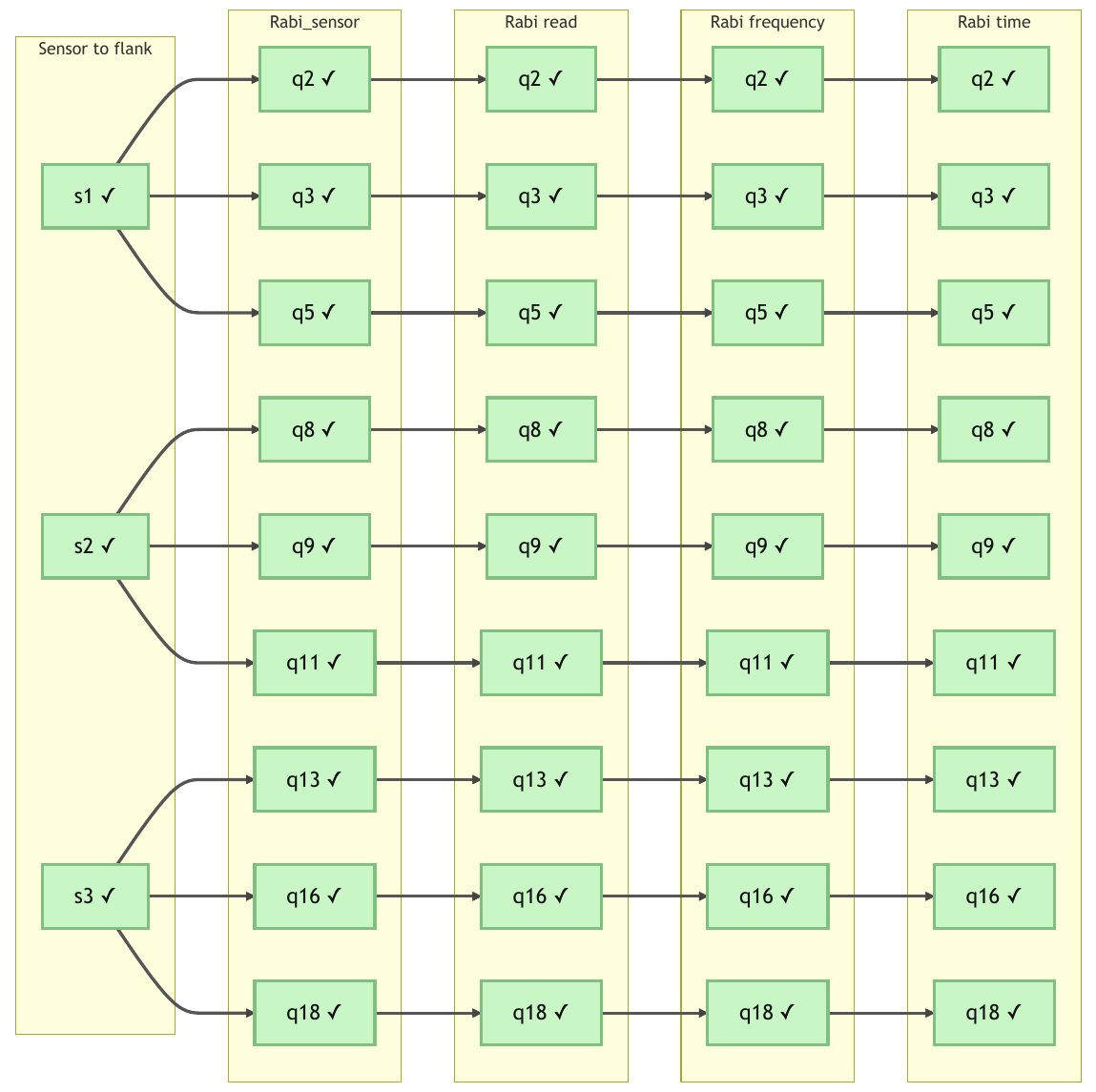}}
\caption{\textbf{Automatic calibration of the qubit array.} Calibration flow for an 18-qubit device showing the sequence of automated routines that are used to maintain stable operation and compensate for daily variations. }
\label{fig:auto_cali}
\end{figure*}

\begin{figure*}[htbp] 
\center{\includegraphics[width=1\linewidth]{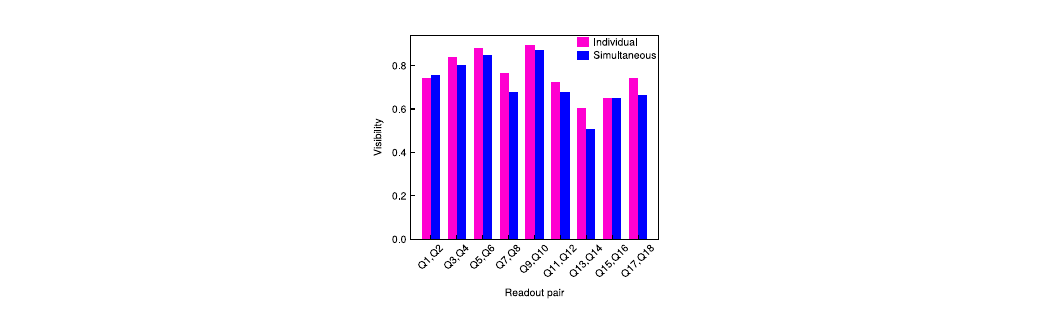}}
\caption{\textbf{Visibility comparison of individual and simultaneous Rabi oscillations.} The visibility is defined as $V=P_\text{max}-P_\text{min}$, where $P_\text{max}$ and $P_\text{min}$ are the maximal and minimal probabilities in a sinusoidal fit to the measured Rabi oscillations. The data used to extract these visibilities is shown in Fig.~\ref{fig:4}b}
\label{fig:vis_compare}
\end{figure*}
\bibliographystyle{naturemag}
\end{document}